\documentclass[letterpaper]{aa}
\usepackage{graphicx}
\usepackage{natbib}
\def\dy{\displaystyle}
\def\h{\hfill\break}

\def\hd{\object{HD\,141569\,A}}
\def\hdb{\object{HD\,141569\,B}}
\def\hdc{\object{HD\,141569\,C}}
\def\bp{\object{$\beta$\,Pictoris}}

\newcommand{\dma}[1]{_{\mathrm{#1}}}
\begin{document}
\title{Structuring the \object{HD\,141569\,A} circumstellar dust disk}
\subtitle{Impact of eccentric bound stellar companions}
\author{J.C. Augereau\inst{1} and J.C.B. Papaloizou\inst{2}}
\institute{ Leiden Observatory, PO Box 9513, 2300 Leiden, The
Netherlands \and Astronomy Unit, School of Mathematical Sciences,
Queen Mary \& Westfield College, Mile End Road, London E1 4NS, UK}
\offprints{J.C. Augereau} \mail{augereau@strw.leidenuniv.nl}
\date{Received; accepted} \titlerunning{Structuring the
\object{HD\,141569\,A} circumstellar dust disk}
\authorrunning{Augereau \& Papaloizou}
\abstract{Scattered light images of the optically thin dust disk
around the 5\,Myr old star \hd\ have revealed its complex asymmetric
structure. We show in this paper that the surface density inferred
from the observations presents similarities with that expected from a
circumprimary disk within a highly eccentric binary system. We assume
that either the two M stars in the close vicinity of \hd\ are bound
companions or at least one of them is an isolated binary companion. We
discuss the resulting interaction with an initially axisymmetric
disk. This scenario accounts for the formation of a spiral structure,
a wide gap in the disk and a broad faint extension outside the
truncation radius of the disk after 10--15 orbital periods with no
need for massive companion(s) in the midst of the disk resolved in
scattered light. The simulations match the observations and the star
age if the perturber is on an elliptic orbit with a periastron
distance of 930\,AU and an eccentricity from 0.7 to 0.9. We find that
the numerical results can be reasonably well reproduced using an
analytical approach proposed to explain the formation of a spiral
structure by secular perturbation of a circumprimary disk by an
external bound companion. We also interpret the redness of the disk in
the visible reported by \citet{cla03} and show that short-lived grains
one order of magnitude smaller than the blow-out size limit are
abundant in the disk. The most probable reason for this is that the
disk sustains high collisional activity. Finally we conclude that
additional processes are required to clear out the disk inside 150\,AU
and that interactions with planetary companions possibly coupled with
the remnant gas disk are likely candidates.
 }
\maketitle
%
\section{Introduction}
Asymmetries and annular structures are common observational features
regardless of the emission process (scattered light or thermal
emission) for the handful of gas-free and optically thin disks
currently resolved around Main-Sequence stars. Surface brightness
maxima peak far from the star from dozens up to about one hundred
AU. Attempts to explain radial and azimuthal structures involve
massive un-resolved planet(s), trapping dust particles into resonances
due to either radial migration of particles sensitive to
Poynting-Robertson drag and/or to radiation pressure
\citep[e.g. \object{$\epsilon$\,Eri}:][]{oze00,qui02} or to outward
migration of the planet \citep[e.g. \object{Fomalhaut} and
\object{Vega}:][]{wya02,wya03}.  Other attempts involve external
stellar companion(s) either bound and observed
\citep[e.g. \object{HR\,4796}:][]{wya99} or unbound and currently
unobserved but having recently approached the close vicinity of the
disk \citep[flyby scenario, e.g. \bp:][]{lar01}. In the edge-on disk
of \bp, the vertical asymmetries are explained by the precession of
planetesimal orbits induced by an inner planetary companion on an
orbit that is inclined to the dust disk \citep[][]{mou97} and the
effect of radiation pressure acting on the smallest grains
\citep[][]{aug01}. But note that the inner planet in this model can be
replaced by any inner misaligned mass distribution with the
appropriate quadrupole moment components and is not dependent on any
particular planet postulate.  An alternative scenario involving dusty
clump formation through stochastic collisions between large
planetesimals has been proposed by \citet{wya02} to explain the
asymmetries noticed in the Fomalhaut disk.

We explore in this paper a source of asymmetry for the optically thin
dust disk surrounding \hd, a B9.5V--A0V star located at about 100\,pc
according to Hipparcos measurements. Coronagraphic images from the
visible to the near-infrared have revealed the complex morphology of
the dusty circumstellar environment of this old Herbig star
\citep{aug99,wei99,mou01,boc03,cla03}.
\begin{figure}
\begin{center}
\vspace*{3cm}
[{\it Figure available in JPEG format or download the
paper at http://www.strw.leidenuniv.nl/$\sim$augereau/newresults.html}]
\vspace*{3cm}
\caption{HST/STIS visible image of the optically thin dust disk around
\hd\ from \citet{mou01}. The two M companions, \hdb\ and C, located in
the North-West region lie outside of the image (see text for precise
location).}
\end{center}
\label{hd141stis}
\end{figure}
One can construct the following sketch of the shape of the dust disk
beyond 100--120\,AU (the edge of the coronagraph) as it appears in
scattered light (see also Figure \ref{hd141stis})~:
\begin{itemize}
\item the disk is composed of two annuli peaked at 200\,AU and 325\,AU
from the star with centers shifted by 20--30\,AU almost along the
minor axis of the disk (East-West direction). The outer ring actually
shows a tightly-wound spiral structure,
\item the two bright predominant annuli  at 200\,AU and 325\,AU have
between them a darker ring or ``gap''. This gap is radially wide
compared to the two annuli,
\item the two bright rings show out of phase brightness asymmetries of
up to factors of 2.5--3 for the outer ring in the visible. These
asymmetries cannot be explained by invoking scattering properties of
the dust grains,
\item an extended diffuse emission is present in the North-East of the
disk and is detected up to more than 600\,AU,
\item the disk brightness sharply decreases between 200 and 150\,AU
rapidly reaching the background level of scattered light images
interior to 150\,AU. This behavior is suggestive of a strong, but
likely not complete, depletion of dust inside 150\,AU,
\item structures with smaller spatial scales are also present such as
a radially thin arc superimposed on the dark lane/gap at a distance of
250\,AU from the star,
\item the maximal vertical optical thickness of the outer ring is
$\sim 2\pm 1\,\times 10^{-2}$ in the visible and the near-infrared.
\end{itemize}
The dust content inside 100--120\,AU remains almost totally
unconstrained despite marginally resolved images in the mid-infrared
\citep{fis00,mar02} which indicate a confirmation of dust depletion
inside $\sim$100\,AU \citep[Fig. 4 from][]{mar02}. A total midplane
optical depth in the visible of $\sim$0.1 has been estimated by
\citet{li03} indicating that the disk is optically thin in all
directions.
\h

\hd\ is not isolated but has two low-mass stellar companions \hdb\ and
\hdc\ located at 7.54'' and 8.93''. Their position angles (PA) are
311.3$\degr$ and 310.0$\degr$ respectively \citep{aug99}. We show in
this paper that the gravitational perturbation of the \hd\ disk by the
detected stellar companions gives a natural explanation for some of
the broadest features observed in scattered light as long as one of
the companions, or both if bound, is on an orbit with high enough
eccentricity.

We detail in section \ref{model} our motivations for exploring the
impact of the observed companions on the shape of disk and we give a
description of the dynamical model we used to address this issue. The
numerical results shown in section \ref{basic} are compared with an
analytic solution to the problem in section \ref{theory}. A surface
density consistent with the resolved images of \hd\ is obtained in
section \ref{param} and we discuss the implications for the dynamics
of the companions. In section \ref{grainsize}, we interpret the
redness of the disk in the visible measured by \citet{cla03} in terms
of minimal grain size in the disk and we discuss the consequences of
these results. We finally point out the limitations of our dynamical
approach in section \ref{discu} and indicate directions for future
work.

\section{Dynamical modeling}
\label{model}
\subsection{Motivations}
We discuss in this paper the possible gravitational influence on the
dust disk of the stars \hdb\ and C observed in the close vicinity of
\hd.  Motivations for exploring the potential gravitational influence
of the detected stellar companions are threefold:
\begin{enumerate}
\item Based on astrometric and radial velocity measurements,
\citet{wei00} argue that B and C, identified as M2 and M4 pre-Main
Sequence stars respectively, have a high probability of forming with
\hd\ a bound triple system. The three stars have similar ages~:
5$\pm$3\,Myr for the two M companions \citep{wei00} and $\sim$5\,Myr
for \hd\ \citep{mer03}. If \hd, B and C are indeed gravitationally
linked, B and/or C are likely to perturb the disk surrounding the
primary and are natural candidates for structuring the disk. In any
case the possible effects of these observed probable companions has to
be explored before invoking other sources of asymmetries in the
disk. Moreover the very latest images of the disk with HST/ACS
strongly support our approach \citep{cla03}.
\item Could undetected companion(s) in the midst of the disk explain
the gap at 250\,AU? \citet{wei99} show that a companion with a mass
larger than 3 times that of Jupiter would have been detected by the
HST/NICMOS instrument but on the other hand an object with a mass 1.3
times that of Jupiter would be sufficient to form and clear the
250\,AU region. While the presence of such companion(s) can not be
ruled out, current planet formation models do not readily account for
the in situ formation of Jupiter-like objects at a distance of a few
hundred AU around a $\sim$5\,Myr old star.
\item Previous modeling attempts to explain the structures in the disk
fail, as noticed by the authors themselves, to reproduce them
\citep{tak01}. The differential migration of dust grains in a
gas disk with an outer sharp edge leads to the formation of a gap and
an outer dust over-density of small particles but this approach, which
apparently requires the presence of an unaccounted for sharp outer
edge to the gas disk, results in a very axisymmetrical structure which
does not match the more recent observations.
\end{enumerate}

\subsection{Model and assumptions}
We consider in this paper the evolution of a collisionless
circumstellar disk of solid particles gravitationally perturbed by an
external bound perturber on an eccentric orbit.

We assume that at least one of HD\,141569\,B and C and possibly both
are bound to \hd. Because the orbital parameters of the two companions
with respect to \hd\ are unknown and because of their close projected
positions we will adopt the simplest assumption that there is a single
perturber assumed to be coplanar with the disk that can interact
strongly with it.  In principle this could be either one of the
companions or a composite of them if they are bound together.  We set
the mass of the single perturber to the total mass of HD\,141569\,B
and C assumed to be respectively M2 and M4 stars. But note
self-similar properties of the simulations enable a scaling to
different companion masses.  This leads to a secondary to primary mass
ratio of $\sim$0.2. The pericenter distance of the perturber is
constant in the model and we vary the eccentricity of its orbit
(Figure \ref{init}). Thus, unless explicitly stated, distances will be
expressed in units of pericentre distance between the perturber and
the primary. Four perturber eccentricities are explored in this
paper~: $e=$ 0.1, 0.3, 0.5 and 0.7. The perturber starts the
simulation at its pericentre position.
\begin{center}
\begin{figure}
\vspace*{2cm}
[{\it Figure available in JPEG format or download the paper at
http://www.strw.leidenuniv.nl/$\sim$augereau/newresults.html}]
\vspace*{2cm}
\caption{A basic sketch of the circumprimary disk at the beginning of
a simulation and of the four perturber orbits considered in the
paper. Distances are expressed in model units (pericentre distance =
1). The orientation of the pericentre, along the $x$ axis in the
direction $ x > 0,$ is fixed throughout the simulations.}
\label{init}
\end{figure}
\end{center}

The initial circumprimary disk consists of $9\times 10^5$ ``massless''
test particles distributed in a 2D axisymmetrical disk from $r=1/6$ to
$r=7/12$ model units with a radial power-law surface density
$\Sigma(r) \propto r^{\alpha}$. The disk is splited up into nine
concentric annuli of $10^5$ particles each with radius following the
relation $r_{i+2}=\left(2 r_{i+1}^{\alpha+2} -
r_{i}^{\alpha+2}\right)^{1/(\alpha+2)}$ ensuring the continuity of the
surface density from one annulus to another. A constant surface
density ($\alpha=0$) is assumed but different initial surface
densities were explored by an {\it a posteriori} processing. This
actually does not impact the general behavior described in this
paper. Particles are initially on circular orbits around \hd.

The orbits of the particles are numerically integrated using a
fifth-order Cash-Karp Runge-Kutta method. Effects on the motion
arising from radiation are neglected. The simulations start with a
time-step of a tenth of the orbital period of the closest particle to
\hd. Then it is adjusted within the code in order to ensure
accuracy. The dynamical response of the disk is numerically followed
over a total span of 169, 116, 70 and 32.5 perturber orbital periods
for the 0.1, 0.3, 0.5 and 0.7 perturber eccentricities
respectively. Every fiftieth of the total span, positions and
velocities of the particles are stored. As an example, a pericentre
distance of 1200\,AU and an eccentricity $e$=0.5 correspond to a total
span of 5\,Myr with a storage process time-scale of $10^5$ years.
%
%
%
\section{Numerical results}
\label{basic}
\begin{figure*}[tbp]
\h
\h
\h
\vspace*{5cm} \h [{\it Figures available in JPEG format. Animations
and paper can be downloaded at
http://www.strw.leidenuniv.nl/$\sim$augereau/newresults.html}]%
\vspace*{7cm}
\caption[]{\label{fig_sigma} Evolution of an initially axisymmetric
circumprimary disk within a binary stellar system for different
perturber eccentricities (see section \ref{model}). The images show
face-on views of the disk surface density in a logarithmic scale. The
upper left image shows the common initial disk with $9\times 10^5$
particles spread over a radial range from $r=1/6$ to $r=7/12$. For
each perturber eccentricity, relevant snapshots have been selected in
order to highlight the dynamical response of the disk to the
gravitational perturbation of the perturber. For each panel, the time
is expressed in perturber orbital periods. The position of pericentre
of the perturber is along the $x$ axis in the right direction (see the
sketch in Figure \ref{init}). The position of the central star is
indicated by the white cross.  Brightness levels are the same from one
image to another allowing direct comparison between all the
snapshots.}
\end{figure*}
Figures \ref{fig_sigma}a to \ref{fig_sigma}d show snapshots of the
temporal evolution of an initially axisymmetrical circumprimary disk
of test particles in an eccentric binary system with a mass ratio of
0.2 for different perturber eccentricities. Several features
characterize the dynamical response of the disk to the perturber among
which the clearing out of the outer disk and the formation of spiral
structures are the most prominent ones.
\begin{figure*}[tbp]
\hbox to \textwidth
{
\parbox{0.34\textwidth}
{
\hspace*{-0.5cm}
\includegraphics[angle=0,origin=tr,width=0.34\textwidth]{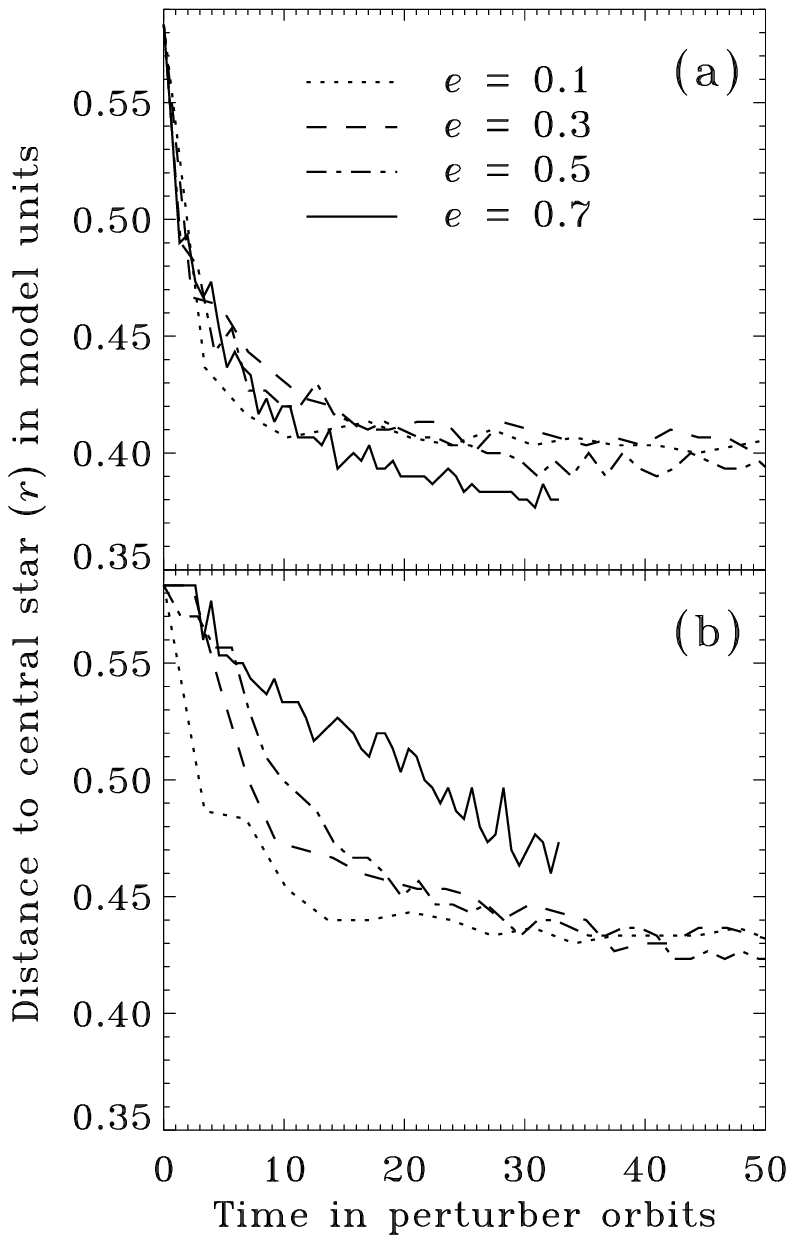}
}
\hspace*{-0.85cm}
\parbox{0.71\textwidth}
{
\includegraphics[angle=0,origin=tr,width=0.71\textwidth]{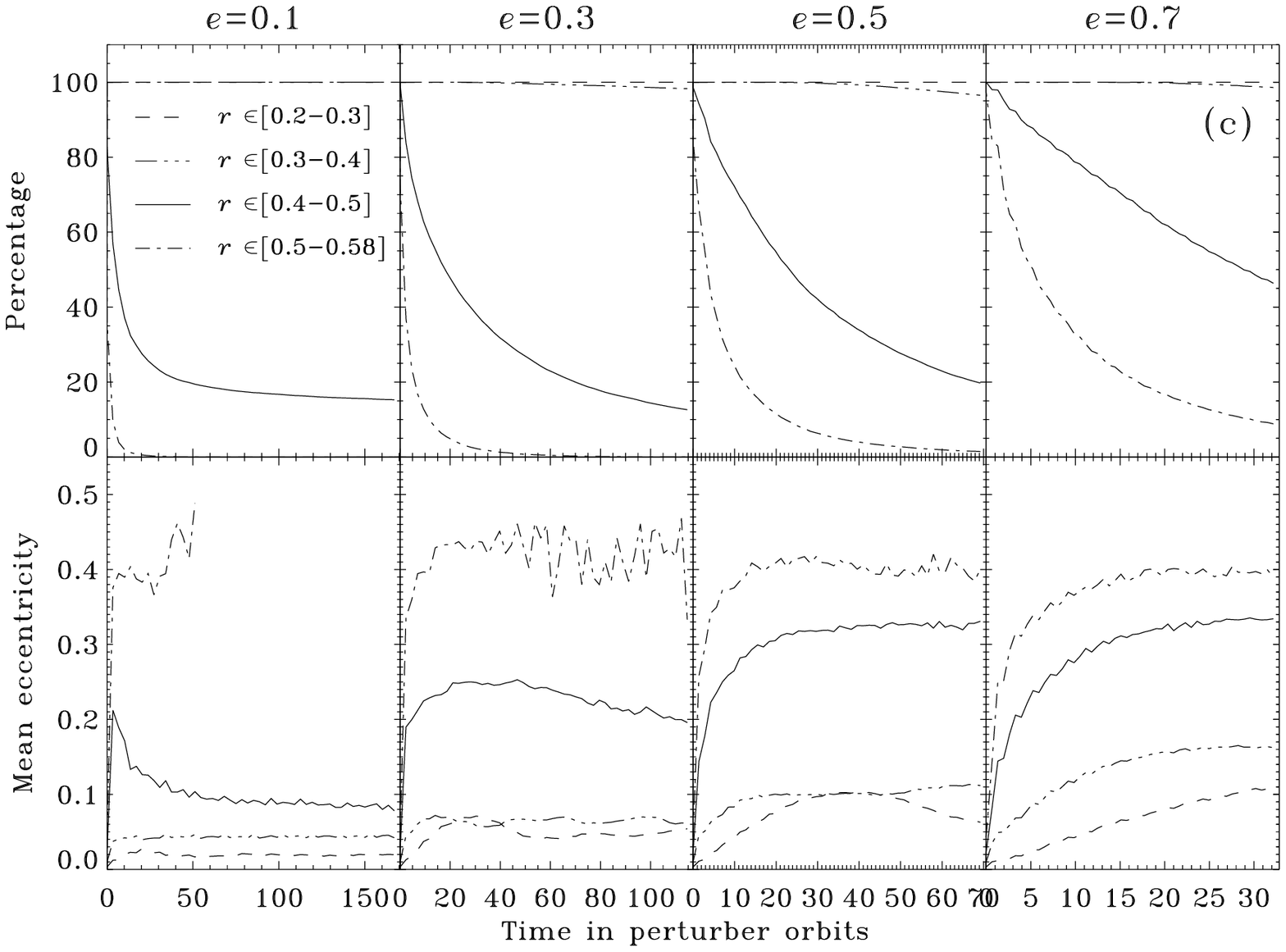}
}
}
\caption[]{\label{fig_pe} {\it Panels} (a) {\it and} (b) : truncation
distance versus time. Distances outside which the surface density is
less than half (resp. a quarter of) the initial surface density are
represented in panel (a) (resp. (b)). Time is expressed in number of
perturber orbital periods. Distances are expressed in model units,
{\it i.e.} normalized to the perturber pericentre distance.  {\it
Panels} (c) : mean eccentricities (lower panels) and fractions (upper
panels) of bound particles versus time for the four different
perturber eccentricities and four ranges of distances to the primary
star (in units of perturber pericentre distance). Time is expressed in
number of perturber orbital periods but the total physical time is the
same for each value of perturber eccentricity ($e$). The range of
distances represented by the solid line on each panel brackets the
truncation distance of the disk.}
\end{figure*}

Particles initially at radii larger than $r=0.4-0.45$ model units
reach highly eccentric orbits after about ten perturber periods
resulting in a gravitational truncation of the disk consistent with
theoretical calculations \citep[e.g.][]{pap77}. The truncation occurs
regardless of the assumed perturber eccentricity $e$ (Figure
\ref{fig_pe}) but low perturber eccentricities increase the efficiency
of particle ejection after a fixed number of orbital periods since the
pericentre distance of the perturber is constant in the model.  For
instance 90\% of the particles initially outside the critical radial
distance of $r=0.45$ model units have been placed on unbound orbits
for $e=0.1$ after 10 perturber revolutions whereas 45\% of them were
still bound to the central star for $e=0.7$ after the same number of
orbits (Figure \ref{fig_pe}).  Together with the low eccentricities
induced by the perturber on the particles interior to $r=0.4$ model
units, this effect results in sharp outer disk edges for low $e$
values. Conversely the disk truncation is less marked for the largest
$e$ values. This is contributed to by the increase of the mean
eccentricity of the bound particles which have the ability to perform
excursions significantly beyond the truncation distance (Figure
\ref{fig_pe}b and c). Actually the disk truncation for large $e$
values is not homogeneous in azimuth as can be noticed in Figure
\ref{fig_sigma} for $e=0.5$ and $e=0.7$.  This is manifest by the
formation of a marked azimuthal asymmetry of the surface density for
distances larger than $r=0.4-0.45$ model units for the largest $e$
values after similar numbers of perturber orbital periods.

Inside the truncation distance, an over-density of particles develops
in conjunction with the truncation process which breaks the
axisymmetry of the initial disk. This spatially coherent over-density
appears close the outer edge of the truncated disk at a distance
$r\sim 0.35$ model units. As shown below, the secular perturbation of
the disk by the perturber is essentially responsible for the formation
of this spiral-shaped structure which is well marked for $e\geq 0.3.$
It progressively winds around the central star becoming radially
thinner with time. The formation of the spiral-shaped over-density of
particles leads in turn to the formation of lower surface density
regions and ultimately results in the opening of a spiral gap. The gap
always appears to be radially wider than the adjacent over-density
spiral.  The contrast between the high and low density regions begins
to increase with time as the spiral propagates inwards. The particle
eccentricity induced by the perturber depends on its eccentricity
(Figure \ref{fig_pe}). This prevents the spiral structure from having
exact self-similar properties as the perturber eccentricity
varies. Nevertheless we notice that the large scale features evolve in
a similar way as a function of the perturber orbital period (Figures
\ref{fig_sigma}b to \ref{fig_sigma}d). The latter characteristic is
used in section \ref{param} in order to derive orbital parameters for
the perturber matching the observations of the HD\,141569 system.
%
%
\section{Spiral structure by secular perturbation of the inner disk}
\label{theory}

We here show that the transient spiral structure in the inner disk
seen in the simulations can be understood as being due to the action
of the time averaged potential due to the perturber.  This accounts
for some of the self-similar scaling properties seen in the
simulations.

We adopt a simple model based on calculating the linear response of a
cold collisionless disk to an orbiting companion in the continuum
limit. In this scheme differential precession would cause the induced
spiral structure to wind up indefinitely generating arbitrarily small
scales and large surface densities.  In a more realistic situation,
the fact that the disk is not a continuum and has a non zero velocity
dispersion will cause the spiral structure to eventually wash out. To
model such effects we introduce an ad hoc decay rate for the spiral
form that increases as its radial scale decreases. Further we impose a
cut off on the magnitude of the surface density calculated using the
linear response. In this way the temporal morphology of the spiral
form apparent in the linear response remains while unphysical effects
arising through the simplicity of the model are suppressed.
\begin{figure*}[tbp]
\begin{center}
\vspace*{2cm}
[{\it Figures available in JPEG format or download the
paper at http://www.strw.leidenuniv.nl/$\sim$augereau/newresults.html}]
\vspace*{2cm}
\caption[]{\label{fig_spiralform} Surface density plots taken at the
number of orbital periods after initiation indicated. These plots are
for a pertuber with $e=0.7$ and $M_p = 0.2M_*$.  The analytical
procedure described in section \ref{theory} was used. The pericentre
distance was set to 1 distance unit and the direction of pericentre is
the same as in Figures \ref{init} and \ref{fig_sigma}. Disk radii
between 1/6 and 0.47 are considered which can thus be directly
compared with Figure \ref{fig_sigma}d. Such a comparison shows that
similar morphology develops at similar times.}
\end{center}
\end{figure*}

We suppose that $(r, \varphi )$ define a cylindrical coordinate system
based on the primary star of mass $M_*$ which exerts a potential $\Psi
= -GM_*/r.$ We take the perturber with mass $M_p$ to be in an
eccentric orbit with semi-major axis $A$ and eccentricity $e.$ The
apsidal line is taken to lie along the $x$ axis ($\varphi = 0$).  The
perturbing potential $\Psi_p$ it produces can be expanded in terms of
a Fourier cosine series in $\varphi$ and time averaged
\citep[eg.][]{ter02}.  It is necessary to retain only the first two
terms in the Fourier series to lowest order in $r$ so that we may
write
\begin{equation}
\Psi_p = \Psi_{p0}  + \Psi_{p1} 
 = -{GM_p r^2\over A^3(1 - e^2)^{3/2}}\left({1\over 4}+{3er\cos\varphi
\over 8A(1-e^2)}\right) \label{secp}.
\end{equation}
The first axisymmetric term $(m=0)$ $,\Psi_{p0} ,$ contributes to the
unperturbed motion in the disk. In particular it leads to a precession
of orbits with small eccentricity at a rate $\omega_p = \Omega -
\kappa,$ with $\Omega$ and $\kappa$ being the angular and epicyclic
frequencies associated with a circular orbit containing the conserved
orbital angular momentum respectively \citep[see eg.][for the basic
definitions and disk dynamics required]{bin87}.  For the time averaged
potential $\Psi +\Psi_{p0},$ assuming, as is the case, that $\omega_p
\ll \Omega,$ one obtains
\begin{equation}
2\Omega\omega_p = {3GM_p \over 2A^3(1 - e^2)^{3/2}}. \label{precec}
\end{equation}
In general this is a function of the radius, $r_0,$ of the circular
orbit. The presence of the second nonaxisymmetric $(m=1)$ term,
$\Psi_{p1}$ in (\ref{secp}) perturbs these to become eccentric.  One
may write
\begin{equation}
r= r_0 + \xi_{r_0} \,\,\, , \,\,\, \varphi = \varphi_0 + \xi_{\varphi_0}/r_0
\end{equation}
where $( \xi_{r_0}, \xi_{\varphi_0})$ denote the components of the
Lagrangian displacement vector in the radial and azimuthal directions
respectively.  Note too that $\varphi_0 = \Omega(r_0)t + \beta_0,$
where $\beta_0$ is a fixed phase associated with and identifying a
particular orbit.

Assuming $ \xi_{r_0} = e_d r_0,$ with $e_d$ being a small disk
eccentricity, in the linear approximation the governing equation is
that of a forced harmonic oscilator taking the form
\begin{equation}
{d^2 \xi_{r_0}\over dt^2} + \kappa^2(r_0) \xi_{r_0} =
-\left({\partial \Psi_{p1}\over \partial r} +
{2\Psi_{p1}\over r}\right)_{r=r_0, \varphi=\varphi_0}.
\label {forco}
\end{equation}
Note that the time derivative is taken along a given unperturbed orbit
and so evaluation of the perturbing potential at $ \varphi=\varphi_0$
introduces a time dependence on such an orbit.
\begin{figure*}[tbp]
%
\vspace*{3cm} [{\it Figures available in JPEG format. Higher
resolution PNG images and paper can be downloaded at
http://www.strw.leidenuniv.nl/$\sim$augereau/newresults.html}]
\vspace*{3cm}
\caption[]{\label{final} {\it Left panel}: dust surface density
derived from HST/ACS images after deprojection, correction for stellar
flux dilution and correction for anisotropic scattering \citep[Figure
7d from][]{cla03}. {\it Middle and right panels}: simulated disk
surface density after $\sim$9 perturber orbital periods. For
illustrative purpose, an argument of pericentre of $\omega=50\degr$
has been chosen. The true anomalies of HD\,141569\,B and C are
indicated on the right panel assuming a disk inclination of 55$\degr$
from pole-on and an observed disk position angle of -3$\degr$. This
leads to deprojected distances for HD\,141569\,B and C of 1078.17\,AU
and 1291.29\,AU respectively.}
\end{figure*}

Equation (\ref{forco}) is easily solved for $\xi_{r_0}$ provided
appropriate boundary conditions are applied. Here we assume that prior
to some time $t=0$ when the companion is introduced the disk is
unperturbed. It is important to emphasize that the formation of
transient spiral structure requires such a 'sudden' introduction of
the perturber. Here 'sudden' means fast compared to the inverse
precession frequency at the disk location of interest. A possibility
would be a distant scattering into an orbit of high eccentricity near
apocentre. A very 'slow' introduction of the perturber through build
up of its mass on a long timescale would not lead to transient spiral
structure.  A solution of (\ref{forco}) corresponding to 'sudden'
introduction in the limit of small $\omega_p/\Omega $ can be expressed
in the form
\begin{equation}
\xi_{r_0} =- {5er_0^2\over 4A(1-e^2)} \left(\cos(\varphi_0)
-\cos(\varphi_0 - \omega_p(r_0) t)\right).
\label{soln}
\end{equation}

The first thing we point out about (\ref{soln}) is that it has a
self-similar scaling property for $e \sim 1.$ Noting that in that
limit $A(1-e^2) \sim 2A(1-e)$ being twice the pericentre distance and
using equation (\ref{precec}), we see that equation(\ref{soln}) gives
the same displacement for a given pericentre distance and perturber
mass ratio after a given number of orbital periods independent of $A.$
This scaling, which is only appropriate for secular perturbations,
occurs in our simulations.

Note too that in the expression (\ref{soln}) the perturber mass only
occurs as linear multiple in $\omega_p$ as given by
equation(\ref{precec}).  Thus different perturber masses produce the
same spiral planform, but on longer timescales for lower mass
perturbers. We have also verified that the simulations show this.
Adjusting either one of the companion masses or their combination can
then be accommodated by scaling the time.

Another aspect indicated above is that because $\omega_p$ depends on
$r_0$ we have differential precession and an unlimited winding up and
shortening of the scale of the disk perturbation.  This occurs through
an increase of the wavenumber $k = (d\omega_p/dr_0) t$ that is implied
in the second term in (\ref{soln}) which is there because of the
'sudden' initial condition.  Such large values of $k$ can in theory
lead to unrealistically large or even singular values for the surface
density.

In fact an arbitrary increase of $k$ does not manifest itself in the
simulations or the physical problem of interest. It does so in the
simplified analysis given above because the disk was assumed to be a
continuum initially with exact circular orbits only.  The introduction
of some velocity dispersion would result in a particle sampling a
range of radii and thus a limit to the size of the value $k$ and the
disk surface density that may occur in any visible perturbation.  Very
short scales, if initiated, would be washed out and decay.  Here we
represent the decay of very short scale disturbances in a simple ad
hoc manner.  We simply reduce the amplitude of the $\cos(\varphi_0 -
\omega_p(r_0) t)$ term in equation (\ref{soln}) by a factor that
depends only on time by replacing it by $\cos(\varphi_0 -
\omega_p(r_0) t) /\sqrt{1+0.025k^2R^3/r_0},$ where $R$ is a fiducial
radius. This reduction causes the second term in (\ref{soln}) and
hence the spiral structure to decay with time. It is clear that the
first and remaining term at large times in (\ref{soln}) does not
produce a spiral form.

We now compare the form of the surface density perturbation resulting
from the above analysis with that found in the simulation.  To do this
we use mass conservation expressed in Lagrangian form, namely
\begin{equation}
\Sigma ={\Sigma_0 r_0\over r |J|},
\end{equation}
where $J = |\partial(r, \varphi)/\partial(r_0, \varphi_0)|$ is the
magnitude Jacobian determinant of the transformation $(r, \varphi )
\rightarrow (r_0, \varphi_0).$ Here $\Sigma$ is the surface density of
the disk under perturbation, while $\Sigma_0$ is the surface density
for the unperturbed disk.  We adopt the relation between $\xi_{r_0}$
and $\xi_{\varphi_0}$ that applies to Keplerian orbits with small
eccentricity, namely $\partial\xi_{\varphi_0} /\partial \varphi_0 =
-2\xi_{r_0}.$ To prevent unphysically large surface densities we
employ a cut off through providing a floor for $J$ which limits the
surface density at $2\Sigma_0.$

We plot the surface density obtained from the above analysis in Figure
\ref{fig_spiralform} for a binary with $e=0.7, M_p = 0.2M_*$ and
pericentre distance $1$ model units. The disk is considered for $1/6 <
r_0 < 0.47$ model units and that corresponds to the section of the
disk illustrated in Figure \ref{fig_sigma}. The number associated with
each plot is the time in orbital periods. Comparison with Figure
\ref{fig_sigma}d for the simulations indicates a similar morphology
regarding the spiral pattern at corresponding times. Note that we have
verified that this morphology is insensitive to the mode of the cut
off at high surface densities or of large radial
gradients. Furthermore because the morphology is generated by the time
averaged secular potential, it does not depend on the precise location
of the perturber on its orbit at any time but only on the direction to
pericentre, here being along the $x$ axis.
%
%
%
%
\section{Application to the \hd\ disk}
\label{param}
We discuss in this section the morphological similarities between the
disk features described in the two previous sections and those
observed in the \hd\ disk. We base our semi-qualitative comparison
between simulations and observations on the dust surface density maps
obtained by \citet[][hereafter C03]{cla03}. These maps have been
derived from HST/ACS deprojected images of the disk seen in scattered
light in B, V and I bands. They take into account the stellar flux
dilution effect assuming the disk is optically thin in all directions
at visible wavelengths (Figure 7c from C03). They take also into
account the anisotropic scattering properties of the grains assuming
forward scattering with an asymmetry factor $g$ in the range
0.25--0.35 consistent with the upper limit assessed by \citet{mou01}
(Figure 7d from C03). Given the relative uncertainties in grain
properties in the \hd\ disk (see nevertheless section
\ref{grainsize}), and thus the uncertainties on grain scattering
properties, we may safely assume in the following that Figures 7c and
7d in C03 bracket a representation of the true surface density of the
dust disk by noting that they respectively correspond to the isotropic
and maximum anisotropic assumptions for scattering.

\begin{figure}[tbp]
\begin{center}
\includegraphics[angle=0,origin=tr,width=\columnwidth]{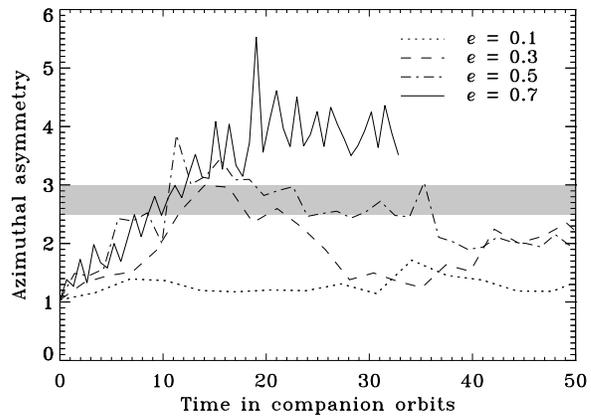}
\caption[]{\label{fig_asym} Azimuthal asymmetry versus time averaged
in a radial region from $r=0.33$ to $r=0.37$ and for different
perturber eccentricities ($e$). The azimuthal asymmetry is here
defined as the ratio of maximal to minimal total numbers of particles
in 30$\degr$ angular sectors. It has been derived from the simulations
presented in Figure \ref{fig_sigma}. The numerical noise leads to a
typical $1\sigma$ uncertainty of 0.1.  The radial range considered
here includes the over-density of particles appearing after $\sim$10
perturber orbital periods and which compares well with the
observations (see section \ref{param}).  The brightness asymmetry
measured in the visible by \citet{cla03} is indicated by the light
grey area. A perturber eccentricity $e\geq 0.3$ and 10--15 perturber
orbital periods are both required in order to reproduce the observed
brightness asymmetry.}
\end{center}
\end{figure}

We interpret the following observational evidences~:
\begin{itemize}
\item the lack of obvious link between the resolved inner and outer
annulus in the \hd\ system,
\item the low surface density of the inner ring compared to the outer
one,
\item the azimuthal asymmetry of the outer ring,
\end{itemize}
as three constraints on the number of perturber orbital periods the
system has evolved for and on the eccentricity of the perturber's
orbit. According to these criteria, well wound structures are ruled
out such as those formed after $\sim$20 perturber orbits.  The
``isotropic'' surface density is better recovered after $\sim$15
perturber orbits as against $\sim$10 perturber orbits for the
``anisotropic'' case (Figures \ref{fig_sigma}b to
\ref{fig_sigma}d). The measured azimuthal asymmetry within the
structure is reproduced for $e\geq 0.3$ (Figure \ref{fig_asym}). For
$e=0.1$, the disk is too poorly contrasted both azimuthally (Figure
\ref{fig_asym}) and radially (Figure \ref{fig_sigma}).

We then conclude that the surface densities of \hd\ inferred from the
HST/ACS observations show large scale features consistent with a
circumprimary disk within an eccentric binary system provided that the
perturber eccentricity is at least larger than 0.3 (most probably 0.5)
and second that the system has been perturbed for a total duration of
$\sim$10--15 orbital periods (Figure \ref{final}). Under such
conditions the spiral-shaped structure newly-formed close to the
truncation distance qualitatively matches the predominant structure
observed at a distance of 325\,AU (also highlighted in red color on
Figure 8a from C03). The position of the over-density which appears
close to $r=0.35$ model units after $\sim$10--15 perturber orbits
(sections \ref{basic} and \ref{theory}) matches the observations if
the pericentre distance is set to $\sim$930\,AU. An argument of
pericentre of $\sim$50$\degr$ (anti-clockwise direction on the
Figures) with an uncertainty of at least $10\degr$ qualitatively
matches the observations. Such orbital parameters for the perturber
are consistent with current deprojected positions of the two
companions assumed to lie in the disk plane and assimilated into a
single perturber in our approach (Figure \ref{final}).

In a direction opposite to that of the over-density of particles, the
simulated disk shows a broad faint extension. A first consequence is
that the outer edge of the disk looks shifted in the direction of this
extension while the central disk hole (artificially introduced in the
simulations) remains centered onto the star. Therefore, the formation
of a spiral structure in the outer regions of the disk as well as the
shift of the center of the outer disk edge provide explanations of the
shifts reported by several authors between the inner and outer
``rings'' \citep{mou01,boc03,cla03}. A second consequence is that the
broad extension opposite to the over-density may account for the
``diffuse emission'' reported by \citet{mou01} in that direction (see
Figure \ref{hd141stis}). An alternative explanation for the ``diffuse
emission'' could be a local enhancement of the collision frequency at
the location of the over-density resulting in the production of a
larger number of small particles with large eccentricities.

Matching the estimated star age of $5\pm3$\,Myr \citep{wei00,mer03}
critically reduces the range of companion eccentricities~: a 2\,Myr
age implies $e=0.78\pm0.03$ while 8\,Myr leads to $e=0.91\pm0.01$. We
made use of the rough self-similar properties applying inside the
truncation distance as noted in sections \ref{basic} and \ref{theory}
in order to derive these values. Lower eccentricities are
theoretically possible but less probable in the framework of a
gas-free disk of solid material unless the perturbation is very recent
(see the discussion in section \ref{theory}).
\section{Minimum grain size}
\label{grainsize}
\citet{cla03} also report the redness of the disk in the visible.  The
disk scatters $25\pm2$\% more stellar flux in the I band than in the B
band and $10\pm0.7$\% more in the I band than in the V
band. Interestingly no color gradient within the disk is measured
which indicates that similar scatterers dominate the scattered light
images at every distance from the star. One can anticipate that the
grains responsible for the color effect are close in size to the
wavelengths discussed here. In this section we further constrain the
size distribution in the disk and discuss some implications.
\subsection{Interpretation of the disk color in the visible}
For simplicity we assume that the differential dust size distribution
$dn(a)$ follows a single power law: $dn(a)\propto a^{-\kappa}da$
between a minimum size $a\dma{min}$ and a maximum size
$a\dma{max}$. We in addition restrict the discussion to compact
grains\footnote{The impact of the porosity is discussed in section
\ref{porosity}}. The \hd\ dust disk shares some similarities with
typical debris disks: it is optically thin and grains are short-lived
\citep{boc03,li03}. A reservoir of mass, presumably an unseen
population of colliding large bodies such as planetesimals, feeds the
disk in observable dust grains. The dust size distribution therefore
extends to large particles that behave as grey scatterers in the
visible and near-IR. The color effect measured in the visible is very
likely produced by the smallest particles of the size distribution
which implies $\kappa > 3$ (mean scattering cross section dominated by
the smallest grains).

In the small range of wavelengths considered here, a noticeable color
index is inconsistent with dominant scatterers in the regime of
geometric optics. Grains larger than about ten times the size $a_0$ at
which the scattering efficiency $Q\dma{sca}$ reaches its maximum
induce fluctuations of $Q\dma{sca}$ of the same order of or smaller
than the uncertainty on the color index reported by C03. This remark
leads to very conservative upper limits on $a\dma{min}$. On the other
hand a strict lower limit on $a\dma{min}$ consistent with the
observations can be derived by considering the grain scattering
behavior in the Rayleigh regime. This approach is detailed in Appendix
\ref{amin} (Eq. \ref{amin_inf}).  Table \ref{tabamin} gives the
results for spherical silicate and graphite compact grains.
\begin{table}[!btp]
\begin{center}
\caption{Constraints on the minimum grain size $a\dma{min}$ in the
disk based on the interpretation of the color indexes measured in the
visible \citep[HST/ACS,][]{cla03}. Compact grains with a differential
size distribution proportional to $a^{-\kappa}$ are assumed. Sizes are
expressed in $\mu$m. They are computed with Eqs. \ref{amin_sup},
\ref{amin_inf} (with $\alpha=3$) and \ref{a0} assuming $\kappa=3.5$
and complex dielectric permitivities of $\epsilon=2.8+\mathrm{i}
\,0.1$ (or $n\simeq 1.67$) for amorphous silicate and
$\epsilon=5.1+\mathrm{i} \,0.6$ (or $n\simeq 2.26$) for graphite
\citep{lao93}. ``B-I'' and ``V-I'' refer to the bands used to measure
the color effects. $a_0$ is computed in the I band.}
\label{tabamin}
\begin{tabular}{rcccc}
 & \multicolumn{2}{c}{lower limit} & & upper limit \\ &
 \multicolumn{2}{c}{on $a\dma{min}$ ($a\dma{min}^{-})$} & & on
 $a\dma{min}$ (10\,$a_0$)\\
& \vspace{-0.3cm}\\
\hline
& \vspace{-0.3cm}\\
 & B-I & V-I & &   \\
& \vspace{-0.4cm}\\
\cline{2-3}
& \vspace{-0.2cm}\\
Silicate : &
0.077 &
0.095 &
& 3.1 \\
 & & & &  \\
Graphite : &
0.041 &
0.051 &
& 1.6 \\
\hline
\end{tabular}
\end{center}
\end{table}

The upper limit on the minimum grain size in the disk must be compared
to the blow-out size $a\dma{pr}$ in the radiative environment of
\hd. \citet{boc03} show that compact spherical grains smaller than
$a\dma{pr}\sim6\,\mu$m are blown away by radiation pressure. We
conclude from our interpretation of the redness of the disk reported
by C03 that the disk contains a large fraction of grains smaller than
the blow-out size limit (so-called $\beta$-meteoroids in the Solar
System). A similar result was derived from the interpretation of a
marginal color effect in the near-infrared \citep{boc03}.
 
According to the results summarized in Table \ref{tabamin}, the size
of the smallest grains in the \hd\ compares to the largest
interstellar grains \citep[e.g.][]{li97, wei01,clay03}, to the
monomers of cometary dust particles \citep[e.g.][]{gre90,kim03} or to
the constituent monomers of interplanetary dust particles, thought to
be of cometary origin, and which exhibit porous aggregate structures
\citep[e.g.][]{bro85}. The interpretation of the scattered light
observations leads to minimum grain sizes consistent with that
proposed by \citet{li03} from their analysis of the thermal emission
of the disk. They show that the spectral energy distribution of the
disk is well reproduced if $0.1 \leq a\dma{min} \leq 1\,\mu$m.
%
%
%
\subsection{Consequences}
The presence of a large amount of small grains, as small as a tenth of
the blow-out size limit, points out on the need for an efficient
mechanism able to continuously replenish the dust disk in fresh
particles, theoretically expelled from the disk after a few orbital
periods, and responsible for the observed color effect. Mean collision
time-scales two or three orders of magnitude smaller than the star age
have been estimated from the vertical optical thickness of the \hd\
disk \citep{boc03}.  Therefore collisional evolution can happen. Given
the grain sizes and the large distances from the star considered here,
Poynting-Robertson drag is not efficient compared to collisions. The
large number of $\beta$-meteoroids in the disk, large enough to
produce a noticeable color effect, probably results from the large
collisional activity of the disk compared to more evolved debris disks
around older stars. According to \citet{kri00} and \citet{aug01}, the
outer disk of \bp\ could also hold a population of
$\beta$-meteoroids. But the shape of the size distribution departs
from a single power law in \citet{kri00}'s model\footnote{Using a
different approach, \citet{the03} also show that dust production by
collisions results in size distributions that significantly depart
from single power laws} and a major consequence is that the
cross-section is dominated by bound particles that have a grey
scattering behavior consistent with the neutral color of the outer
\bp\ disk in the visible \citep[e.g.][]{lec93}.

Alternatively, we can further consider the optical behavior of the
grains. The scattering properties of an aggregate might be similar to
that of the constituent particles if the aggregate is porous enough.
Could the dominant scatterers responsible for the images of the \hd\
disk be monomers packed into porous bound grains? We let $P$ the
porosity of an aggregate and $a\dma{pr}(P)$ the corresponding blow-out
size limit. For large $P$ values, the number of monomers required to
form a grain of size $a\dma{pr}(P)$ can be approximated by
\begin{eqnarray*}
N \sim (1-P) \left(\frac{a\dma{pr}(P)}{a\dma{m}} \right)^3
\sim (1-P)^{-2} \left(\frac{a\dma{pr}(P=0)}{a\dma{m}} \right)^3
\end{eqnarray*}
where $a\dma{m}$ is the mean size of the monomers. For \hd, the ratio
$a\dma{pr}(P=0)/a\dma{m}$ is of the order of 10. If the scattered
light images are indeed tracing monomers in very porous bound
aggregates, then the smallest grains in the \hd\ system are made of at
least a few thousands of monomers similar in size to the constituents
of the interplanetary dust particles. The major problem with this
alternative scenario is that collisional time-scales are very likely
short enough to efficiently split up such large aggregates into their
constituent monomers producing an unavoidable population of particles
small compared to the blow-out limit.
\section{Conclusion}
\label{discu}
We have explored in this paper the possible impact of the two M
companions of \hd\ on the dynamics of its circumprimary dust disk.
Provided that at least one of the two companions is bound to \hd\ and
on an orbit of high eccentricity, the tightly wound and asymmetric
spiral structure at $\sim 325\,$AU can be reproduced. By contrast, the
disk interior to the spiral structure looks depleted as observed. The
inner ring which has a low surface density compared to the outer
structure is not reproduced in this approach. From the redness of the
disk in the visible we deduce that the disk contains a large fraction
of short-lived grains small compared to the blow-out size limit and
implying high collisional activity.

How the contrasting structures of the \hd\ disk are affected by
collisional activity and the effects of radiation pressure are key
issues that will need to be explored. The dynamics of the disk is made
even more complex by the fact that gas drag may also play a role. The
system indeed contains a remnant amount of gas that was detected for
the first time by \citet{zuc95}. Whether the gas disk is extended and
massive enough to impact the dynamics and the shape of the dust disk
especially at very large distances (around 300-350\,AU) is an issue
that will be addressed in a near future thanks to recently resolved
images of the CO disk \citep{aug03}.

Our model does not provide an explanation of the observed depletion of
solid material inside 150\,AU. An additional process is required to
clear out the inner disk and to produce the sharp edge observed around
150\,AU. Together with the surprising detection of H$_3^{+}$ in the
\hd\ disk by \citet{bri02}, previously observed in atmospheres of the
giant Solar planets, these observations are suggestive clues for
speculating on the presence of planetary companions in the inner disk.
The coupling of putative newly-formed massive companions in the inner
disk with the gas disk resolved by \citet{aug03} is an exciting
prospect. The disk-planet coupling has been theoretically addressed by
several authors and the orbital parameters of some of the detected
extra-solar planets are understood as the final result of this
interaction. The \hd\ system is young enough (only a few Myr) with a
still reasonably large amount of remnant gas so that disk-planet
coupling might still act. Future observations of the \hd\ system
should judiciously focus on the inner regions to better assess how
empty and perhaps structured the inner gas and dust disk is, and on
opportunities for detecting possible sub-stellar objects inside
150\,AU.
%
\begin{acknowledgements}
We thank A.M. Lagrange and D. Mouillet for helpful discussions on the
interpretation of the high resolution images of the \hd\ disk. We also
thank M. Wyatt for carefully refereeing the paper, P. Th\'ebault for
useful comments and the ACS Science Team for kindly providing HST/ACS
images. J.C.A. is deeply grateful to E.F. van Dishoeck for her
permanent support in the course of this work at Leiden Observatory and
to C. Eiroa for the discussions on the distance of \hd. J.C.A. was
supported by a fellowship from the European Research Training Network
``The Origin of Planetary Systems'' (PLANETS, contract number
HPRN-CT-2002-00308) at Leiden Observatory.

\end{acknowledgements}
%
%
%
%
\appendix
\section{Interpretation of a scattering color effect in terms
of minimum grain size}
\label{amin}
The Mie scattering efficiency $Q\dma{sca}$ of spherical and
homogeneous dust particles only depends on the real part
$n_{\lambda}=Re(m_{\lambda})$ of the complex refraction index of the
grain $m_{\lambda}=\sqrt{\epsilon_{\lambda}}$ and on the so-called
size-parameter $x=2\pi a/\lambda$ where $a$ is the grain radius and
$\lambda$ the wavelength. $Q\dma{sca}$ follows three basic well-known
regimes \citep{boh83}~:
\begin{enumerate}
\item for small $x$ values ($a\ll\lambda$, Rayleigh regime),
$Q\dma{sca} \propto x^4$.
\item for large $x$ values ($a\gg\lambda$) $Q\dma{sca} \sim constant$
(grey scattering).
\item otherwise incident rays and rays passing through the grain
produce interferences resulting in extinction maxima (then scattering
maxima) peaking at $x_i=(2i+1)\pi/2(n_{\lambda}-1)$ where $i$ is a
positive integer or zero.  For a given wavelength, the first peak
appears for grains of size
\begin{eqnarray}
\label{a0}
a_0=\frac{\lambda}{4(n_{\lambda}-1)} \,\,\,\,.
\end{eqnarray}
We note
\begin{eqnarray}
\label{x0}
x_0=2\pi a_0 / \lambda = \frac{\pi}{2(n_{\lambda}-1)} \,\,\,\,.
\end{eqnarray}
\end{enumerate}
\h
We propose an approach allowing the interpretation of a measured
scattering color effect in the context of an optically thin
circumstellar disk. The color effect is expressed by
\begin{eqnarray*}
C=\frac{\Phi_1/\Phi^*_1}{\Phi_2/\Phi^*_2}
\end{eqnarray*}
where $\Phi_1$ and $\Phi_2$ are the measured circumstellar flux at
wavelengths $\lambda_1$ and $\lambda_2 > \lambda_1$; $\Phi^*_1$ and
$\Phi^*_2$ are the stellar flux at $\lambda_1$ and $\lambda_2$. We
assume in the following that the grains have similar complex optical
indices in the range of wavelengths considered. We note
$n=n_{\lambda}$ and $m=m_{\lambda}$. Anisotropic scattering properties
are then similar and the scattering efficiency $Q\dma{sca}$ only
depends on the size parameter $x=2\pi a/\lambda$ for spheres of radius
$a$. In the optically thin regime the color effect can be then written
\begin{eqnarray*}
C = \frac{\langle \sigma_1 \rangle}{\langle \sigma_2 \rangle } =
\frac{\int_{a\dma{min}}^{a\dma{max}} Q\dma{sca}(\frac{2\pi
a}{\lambda_1}) a^{2-\kappa} da}{\int_{a\dma{min}}^{a\dma{max}}
Q\dma{sca}(\frac{2\pi a}{\lambda_2}) a^{2-\kappa} da}
\end{eqnarray*}
where $\langle \sigma_1 \rangle$ and $\langle \sigma_2 \rangle$ are
the scattering cross-sections averaged over the grain size
distribution $dn(a)\propto a^{-\kappa}da$. $a\dma{min}$ is the minimum
grain size and $a\dma{max}$ is the maximum grain size. For the sake of
simplicity we assume $3<\kappa<7$ . We also assume that $a\dma{max}\gg
a_0$ at both wavelengths $\lambda_1$ and $\lambda_2$. Thus, the color
effect does not depend on $a\dma{max}$ and this allows us to replace
$a\dma{max}$ by $+\infty$ in the following.  The color effect $C$ can
be rewritten
\begin{eqnarray*}
C = \frac
{\int_{x_1}^{+\infty} Q\dma{sca}(x)
\lambda_1^{3-\kappa} x^{2-\kappa} dx}
{\int_{x_2}^{+\infty} Q\dma{sca}(x)
\lambda_2^{3-\kappa} x^{2-\kappa} dx}
\end{eqnarray*}
where
\begin{eqnarray*}
x_1 = \frac{2\pi a\dma{min}}{\lambda_1} \,\,\,\,\,\mathrm{and}\,\,\,\,\, x_2 =
\frac{2\pi a\dma{min}}{\lambda_2}\,\,\,.
\end{eqnarray*}
Since $x_1 > x_2$ we rewrite $C$ in the form
\begin{eqnarray}
\label{eq1}
C = %
\left(\frac{\lambda_2}{\lambda_1}\right)^{\kappa-3} \times 
\left(1- \frac{\int_{x_2}^{x_1} Q\dma{sca}(x) x^{2-\kappa} dx}
{\int_{x_2}^{+\infty} Q\dma{sca}(x) x^{2-\kappa} dx}\right)
\,\,\,\, .
\end{eqnarray}
\subsection{Upper limit on $a\dma{min}$ in the Rayleigh regime}
We now consider minimum grain sizes $a\dma{min}$ such that the
scattering efficiencies $Q\dma{sca}(x_1)$ and $Q\dma{sca}(x_2)$ fall
into the Rayleigh regime. The basic idea is to compute an upper limit
of the minimum grain size in the circumstellar disk consistent with
the observed color effect in the regime of grains small compared to
the wavelength. This gives
\begin{eqnarray}
\label{eq2}
\int_{x_2}^{x_1} Q\dma{sca}(x) x^{2-\kappa} dx = 
\frac{Q\dma{sca}(x_2)} {x_2^{4} (7-\kappa)} \left[
\left(\frac{\lambda_2}{\lambda_1}\right)^{7-\kappa} - 1 \right]
x_2^{7-\kappa}
\,\,\,\, .
\end{eqnarray}
We compute the integral between $x_2$ and $+\infty$ in equation
\ref{eq1} in two steps. Between $x_2$ and $x_0$ (given by equation
\ref{x0}) the Rayleigh regime is accurately obeyed except for $x$
values close to $x_0$ where the Rayleigh law over-estimates
$Q\dma{sca}$. Therefore
\begin{eqnarray}
\label{eq3}
\int_{x_2}^{x_0} Q\dma{sca}(x) x^{2-\kappa} dx \leq
\frac{Q\dma{sca}(x_2)} {x_2^{4} (7-\kappa)}
\left(x_0^{7-\kappa} - x_2^{7-\kappa} \right)
\,\,\,\, .
\end{eqnarray}
The scattering efficiency reaches its maximum value for $x=x_0$.
Assuming that $Q\dma{sca}$ is a constant for $x$ larger than $x_0$
leads to an upper limit of the integral between $x_0$ and $+\infty$
\begin{eqnarray}
\label{eq4}
\int_{x_0}^{+\infty} Q\dma{sca}(x) x^{2-\kappa} dx \leq
Q\dma{sca}(x_0) \frac{x_0^{3-\kappa}}{\kappa-3}
\end{eqnarray}
with
\begin{eqnarray}
\label{eq5}
Q\dma{sca}(x_0) \leq Q\dma{sca}(x_2)\left(\frac{x_0}{x_2}\right)^4
\,\,\,\, .
\end{eqnarray}
By combining equations \ref{eq1} and \ref{eq2} and inequalities
\ref{eq3} to \ref{eq5} we finally obtain
\begin{eqnarray*}
x_2 \leq x_0 \left(
\frac{1+
\frac{\dy \left(\lambda_2/\lambda_1\right)^{7-\kappa}-1}
{\dy 1-C\left(\lambda_2/\lambda_1\right)^{3-\kappa}}}
{\dy 1+\left(
\frac{\dy 7-\kappa}{\kappa-3}\right)}
\right)^{1/(\kappa-7)} \,\,.
\end{eqnarray*}
This leads to $a\dma{min} \leq a\dma{min}^+$ where
\begin{eqnarray}
\label{amin_sup}
a\dma{min}^+ = \frac{\lambda_2}{4(n-1)}
\left(
\frac{1+
\frac{\dy \left(\lambda_2/\lambda_1\right)^{7-\kappa}-1}
{\dy 1-C\left(\lambda_2/\lambda_1\right)^{3-\kappa}}}
{\dy 1+\left(
\frac{\dy 7-\kappa}{\kappa-3}\right)}
\right)^{1/(\kappa-7)}
\,.
\end{eqnarray}
Minimum grain sizes $a\dma{min}$ obeying to the Rayleigh approximation
and consistent with the color measurement $C$ must then be smaller
than $a\dma{min}^+$. We recall that the former equation does ensure
that such grains do exist in the disk. Larger grains, also consistent
with the color index but that do not follow the Rayleigh approximation
are not excluded.
\subsection{Lower limit on $a\dma{min}$ in the Rayleigh regime}
A lower limit on the minimum grain size in the Rayleigh regime and
consistent with the observations can similarly be derived provided
that a real $\alpha$ fulfilling the following criteria can be found~:
\begin{enumerate}
\item $1 \leq \alpha \leq x_0/x_2$,
\item $Q\dma{sca}(x_0/\alpha) = Q\dma{sca}(x_2)/x_2^4 \,\times\,
\left(x_0/\alpha\right)^4$, \textit{i.e.} $Q\dma{sca}(x)$ obeys the
Rayleigh approximation for $x$ from $x_2$ to $x_0/\alpha$,
\item $Q\dma{sca}(x_0/\alpha) \leq 1$, \textit{i.e.} the scattering
efficiency for $x=x_0/\alpha$ is smaller than the value of
$Q\dma{sca}(x)$ in the geometric optics regime.
\end{enumerate}
Then 
\begin{eqnarray}
\label{eq7}
\int_{x_2}^{x_0/\alpha} Q\dma{sca}(x) x^{2-\kappa} dx =
\frac{Q\dma{sca}(x_2)} {x_2^{4} (7-\kappa)}
\left[\left(\frac{x_0}{\alpha}\right)^{7-\kappa} - x_2^{7-\kappa} \right]
\end{eqnarray}
and
\begin{eqnarray*}
\int_{x_0/\alpha}^{+\infty} Q\dma{sca}(x) x^{2-\kappa} dx \geq
\int_{x_0/\alpha}^{+\infty} Q\dma{sca}\left(\frac{x_0}{\alpha}\right)
x^{2-\kappa} dx
\end{eqnarray*}
or
\begin{eqnarray}
\label{eq8}
\int_{x_0/\alpha}^{+\infty} Q\dma{sca}(x) x^{2-\kappa} dx \geq
\frac{Q\dma{sca}(x_2)} {x_2^{4} (\kappa-3)}
\left(\frac{x_0}{\alpha}\right)^{7-\kappa}
\,\,\,\, .
\end{eqnarray}
Inequality \ref{eq8} together with equations \ref{eq1}, \ref{eq7},
\ref{x0} and \ref{amin_sup} give
\begin{eqnarray}
\label{amin_inf}
a\dma{min} \geq a\dma{min}^- \,\,\, \mathrm{where} \,\,\, a\dma{min}^-
= a\dma{min}^+ / \alpha
\,\,\,\, .
\end{eqnarray}
We add a few useful remarks about the criteria defining $\alpha$. The
first criterion must be used to check the self-consistency of the
result. The second criterion can consider the fact that size
parameters a few times smaller than $x_0$ already ensures that the
scattering efficiency follows the Rayleigh law with a good
approximation. In order to fulfill the third criterion one should use
the precise Rayleigh approximation \citep[e.g.][]{boh83}
\begin{eqnarray*}
Q\dma{sca}(x) = \frac{8}{3}x^4
\left|\frac{\epsilon-1}{\epsilon+2} \right|^2
\,\,\,\, .
\end{eqnarray*}
Solving the previous equation for $x=x_0/\alpha$ and $Q\dma{sca}(x)=1$
gives an upper limit an $\alpha$. If grains are very porous and made
of inclusions small compared to the wavelength, then $\epsilon$ should
be replaced by $\epsilon\dma{eff}$ where $\epsilon\dma{eff}$ is given
by equation \ref{eps} and $x_0$ is given by equation \ref{x0P}.
\subsection{Grain porosity}
\label{porosity}
Equations \ref{amin_sup} and \ref{amin_inf} which depend on $a_0$ (or
$x_0$) have been obtained by assuming that grains are spherical and
compact.  In the following, we assess the impact of the grain porosity
$P$ on $a_0$ (or $x_0$). A way to go towards the solution is to use an
effective medium theory as described in details in \citet{boh83}. Here
we consider the Maxwell-Garnett theory. In that framework, a porous
grain made of a single material is likened to a two component grain
made of a matrix of vacuum and inclusions of solid material. It is
important to note that this theory is only valid for inclusions
(considered as inhomogeneities into the matrix) small compared to the
wavelength. If these conditions are fulfilled then the effective
dielectric permettivity of the porous grain is then given by
\begin{eqnarray}
\label{eps}
\epsilon\dma{eff} =
1+\frac{3(1-P)(\epsilon-1)/(\epsilon+2)}{1-(1-P)(\epsilon-1)/(\epsilon+2)}
\end{eqnarray}
where $\epsilon=\epsilon\dma{r}+\mathrm{i} \epsilon\dma{i}$ is the
complex dielectric permettivity of the solid material (inclusions in
the Maxwell-Garnett approach).

The real part of $\epsilon\dma{eff}$ can be approximated by
$n_{\lambda}\simeq \sqrt{Re(\epsilon\dma{eff})}$ provided that
$Im(\epsilon\dma{eff})^2 \ll Re(\epsilon\dma{eff})^2$. If the latter
condition is fulfilled, then the Taylor expansion of
$(n_\lambda-1)/(1-P)$ with respect to the porosity $P$ gives
\begin{eqnarray*}
\frac{n_\lambda-1}{1-P} = \left(\sqrt{\epsilon\dma{r}}-1\right) +
\rho_1 P + \textrm{O}(P^2)
\end{eqnarray*}
with $0\leq P<1$ and
\begin{eqnarray*}
\rho_1 = \frac{5\epsilon\dma{r}-
6\sqrt{\epsilon\dma{r}}+\epsilon\dma{i}^2
-\epsilon\dma{r}^2+2}{6\sqrt{\epsilon\dma{r}}}\,\,\, .
\end{eqnarray*}
Therefore, the grain size for which $Q\dma{sca}$ reaches a first
maximum is
\begin{eqnarray}
\label{a0P}
a_0=\frac{\lambda}{4(n_{\lambda}-1)} \simeq
\frac{\lambda (1-P)^{-1}}{4\left(\sqrt{\epsilon\dma{r}}-1 + 
\rho_1 P\right) }\,\,\, .
\end{eqnarray}
The corresponding size parameter is
\begin{eqnarray}
\label{x0P}
x_0 \simeq
\frac{\pi (1-P)^{-1}}{2\left(\sqrt{\epsilon\dma{r}}-1 + 
\rho_1 P\right) }\,\,\, .
\end{eqnarray}
%
%
%
{}
\clearpage

\end{document}